\documentclass[prb,twocolumn,a4paper,showpacs,floatfix,superscriptaddress]{revtex4}  

\usepackage{graphicx}
\usepackage{amsmath}

\graphicspath{{./IMAGES/}}

\begin{document}
\title{A class of exactly solvable models for the Schr\"{o}dinger equation}

\author{C. A. Downing}
\email[]{c.a.downing@exeter.ac.uk}
\affiliation{School of Physics, University of Exeter, Stocker Road, Exeter EX4 4QL, United Kingdom}

\date{\today}

\begin{abstract}
We present a class of confining potentials which allow one to reduce the one-dimensional Schr\"{o}dinger equation to a named equation of mathematical physics, namely either Bessel's or Whittaker's differential equation. In all cases, we provide closed form expressions for both the symmetric and antisymmetric wavefunction solutions, each along with an associated transcendental equation for allowed eigenvalues. The class of potentials considered contains an example of both cusp-like single wells and a double-well.
\end{abstract}

\pacs{03.65.Ge, 03.65.Fd, 31.15.-p}
\maketitle

\section{\label{intro}Introduction}

Exact solutions of the steady-state one-dimensional Schr\"{o}dinger equation\cite{Schrodinger} for a particle of mass $m$, energy $E$, and with an external potential $V(x)$,
\begin{equation}
\label{SE}
\frac{-\hbar^2}{2m} \frac{d^2}{dx^2} \psi(x) + V(x) \psi(x) = E \psi(x),
\end{equation}
are not only of a purely mathematical interest or useful as a testbed of numerical, perturbative or semi-classical treatments, but are important to elucidate interesting physics at an analytic level of realistic systems. 

Since the historic solutions\cite{Morse, Eckart, Rosen, Teller, Manning} found at the advent of quantum mechanics there has been much effort in the community to find further exact solutions,\cite{Scarf, Loudon, Whitehead, Pertsch, Wang, Parfitt, Bagrov} either by use of special functions, or via the ideas of the factorization method\cite{Infeld} and supersymmetric quantum mechanics.\cite{Cooper, Mallow} Lately, finding quasi-exact solutions,\cite{Turbiner, Ushveridze, Bender, DowningHeun, Hartmann} where only some of the eigenfunctions and eigenvalues are found explicitly, has also become a popular pursuit. 

\begin{figure}[htbp] 
\includegraphics[width=0.5\textwidth]{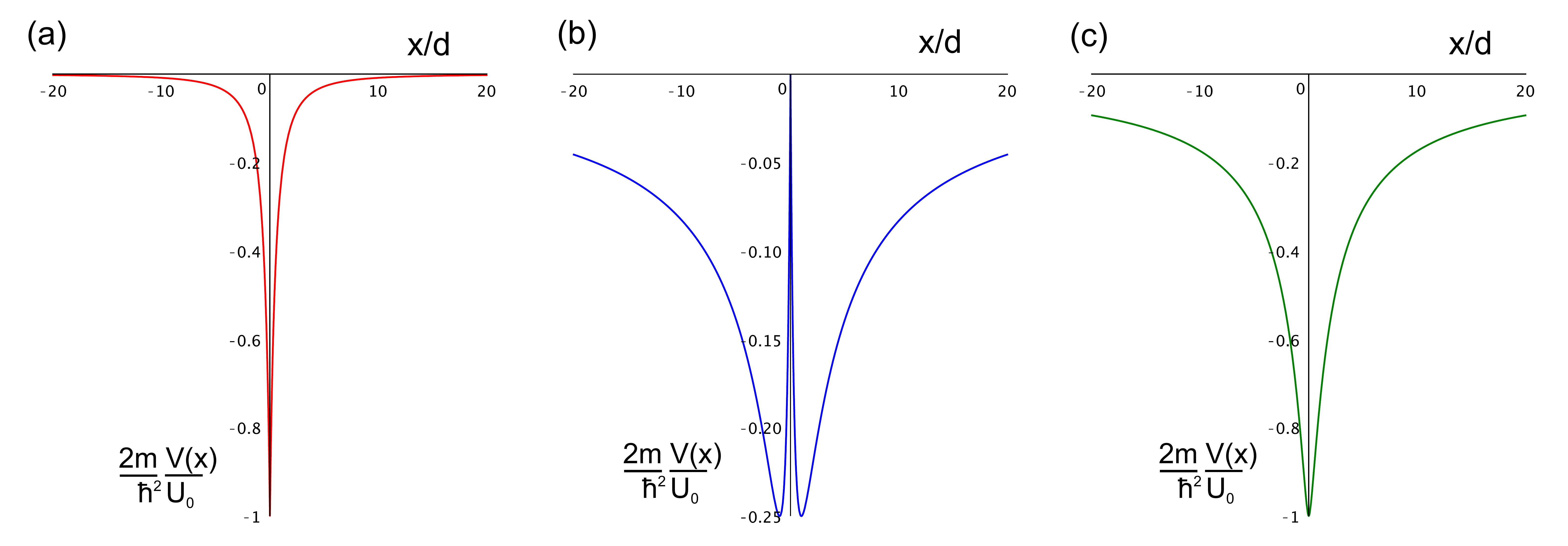}  
\caption{The class of potentials considered,  Eq.~\eqref{potential}, with (a) $p=0$, (b) $p=1$ and (c) $p=2$. Please note in (c) we have shifted the energy reference level and redefined constants, as in Eq.~\eqref{p20}. }
\label{fig:class}
\end{figure}

Here we investigate the following class of attractive confining potential
\begin{equation}
\label{potential}
V(x) = -\frac{\hbar^2}{2m} U_0 \frac{ (|x|/d)^p}{(1+|x|/d)^2}, \qquad p = 0, 1, 2
\end{equation}
with the parameters $U_0>0$ and $d>0$ describing the well depth and width respectively, and the class parameter $p$ defining either: a steep well falling as $1/|x|^2$ $(p=0)$; a double-well $(p=1)$; or a shallow well dropping as $1/|x|$ $(p=2)$, as we plot in Fig.~\ref{fig:class}.

We have also investigated the cases $p=-1, -2$ and found, whilst a similar change of variable to what is used here leads to a confluent Heun differential equation,\cite{Heun} attempts at finding closed form solutions are frustrated by the stringent conditions required for confluent Heun polynomials.\cite{Ronveaux} 

Whilst the importance of single wells in physics and chemistry is well known, the double-well problem is equally interesting, with applications in areas including double-well tunneling,\cite{Gillan} semiconductor heterostructures,\cite{Alferov} atom transfer in a scanning tunneling microscope,\cite{Budau} Bose-Einstein condensation\cite{Spekkens} and instantons.\cite{Coleman}

The rest of this work details our search for bound state $(E < 0)$ solutions of Eq.~\eqref{SE} with the family of potentials Eq.~\eqref{potential}, namely we solve
\begin{equation}
\label{SE2}
\psi''(x) + \left( U_0 \frac{ (|x|/d)^p}{(1+|x|/d)^2} -\kappa^2 \right) \psi(x) = 0,
\end{equation}
where $'$ denotes taking a derivative with respect to $x$ and $\kappa^2 = -2 m E /\hbar^2$, for the class parameter $p=0,1,2$ in Sec.~\ref{p=0}, Sec.~\ref{p=1} and Sec.~\ref{p=2} respectively. We draw some conclusions in Sec.~\ref{conc}.

\section{\label{p=0}Steep single well: \lowercase{p}=0}

\begin{figure}[htbp] 
\includegraphics[width=0.45\textwidth]{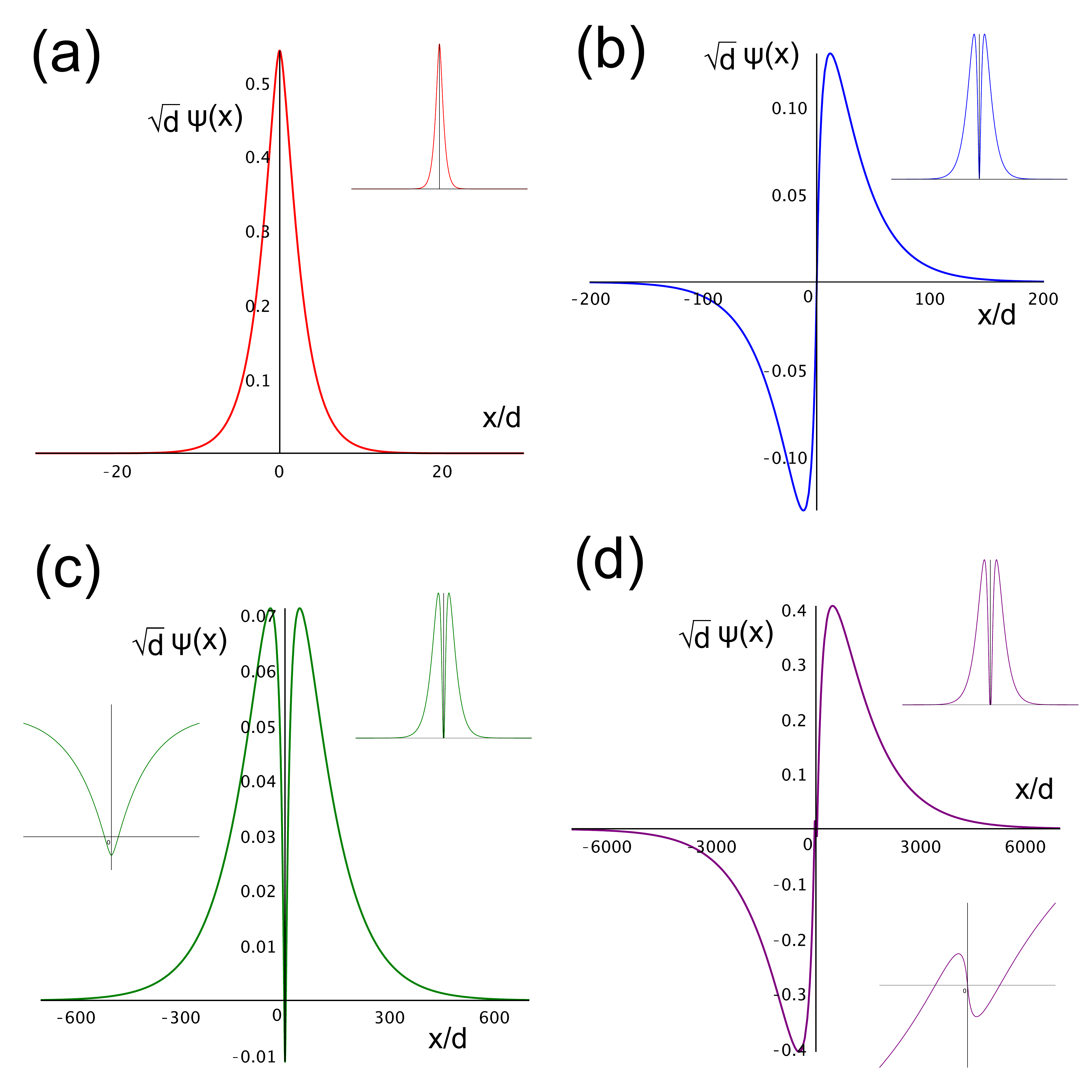}  
\caption{The four lowest eigenstates of the  $p=0$ potential, with $U_0 d^2 = 1$ and (a) $\kappa d = 0.477$, (b) $\kappa d = 0.0373$, (c) $\kappa d = 0.0111$, and (d) $\kappa d = 0.000911$. Inset: probability density. In (c) and (d) a zoom in close to the origin shows the number of nodes more clearly.}
\label{fig:p0plots}
\end{figure}

We substitute $p=0$ into Eq.~\eqref{SE2} and find with the change of variables $\xi = \kappa d (1 \pm x/d)$, where $(+)$ and $(-)$ are taken in half-axes $x>0$ and $x<0$ respectively, the following Schr\"{o}dinger equation
\begin{equation}
\label{p01}
\psi''(\xi) + \left( \frac{U_0 d^2}{\xi^2} - 1 \right) \psi(\xi) = 0.
\end{equation}
Seeking a solution in the form $\psi(\xi) = \xi^{1/2} f(\xi)$ yields the modified Bessel differential equation\cite{Abramowitz} 
\begin{equation}
\label{p02}
\xi^2 f''(\xi) + \xi f'(\xi) - \left( \xi^2 + \alpha^2 \right) f(\xi) = 0,
\end{equation}
with order $\alpha = \sqrt{\frac{1}{4} - U_0d^2}$. The asymptotically decaying solution is the modified Bessel function of the second kind, defined through the modified Bessel function of the first kind $I_{\alpha} (\xi) = \sum_{m=0}^{\infty} \frac{1}{m!} \frac{1}{\Gamma(m+\alpha+1)} \left( \frac{\xi}{2}\right)^{2m+\alpha}$ as follows\cite{Abramowitz} 
\begin{equation}
\label{p03}
 K_{\alpha} (\xi) = \frac{\pi}{2} \frac{I_{-\alpha} (\xi) - I_{\alpha} (\xi)}{\sin(\alpha \pi)},
\end{equation}
which has the large $|\xi|$ asymptotic expansion
\begin{equation}
\label{p03b}
 K_{\alpha} (\xi) \sim \sqrt{\frac{\pi}{2 \xi}} e^{-\xi}.
\end{equation}

The full solution should include two constants $c_{\pm}$ arising from the regions $x \ge 0$ and $x \le 0$ respectively
\begin{equation}
\label{p03c}
 \psi_{\pm}(\xi) =  \frac{c_{\pm}}{\sqrt{d}} \xi^{1/2}  K_{\alpha} (\xi),
\end{equation}
and, upon matching these solutions and their first spatial derivatives across the boundary $x=0$, we find the odd solutions have eigenvalues defined by the transcendental equation
\begin{equation}
\label{p04}
 K_{\alpha} (\kappa d) = 0,
\end{equation}
along with the condition $c_{+} = - c_{-}$, whilst the even solutions have eigenvalues given by
\begin{equation}
\label{p05}
 \frac{K_{\alpha+1} (\kappa d)}{K_{\alpha} (\kappa d)} = \frac{\alpha + 1/2}{\kappa d},
\end{equation}
as well as the restriction $c_{+} = c_{-}$. The form of these conditions, Eq.~\eqref{p04} and Eq.~\eqref{p05}, are familiar from the pioneering work of Loudon\cite{Loudon} on the one-dimensional hydrogen atom.

\begin{figure}[htbp] 
\includegraphics[width=0.45\textwidth]{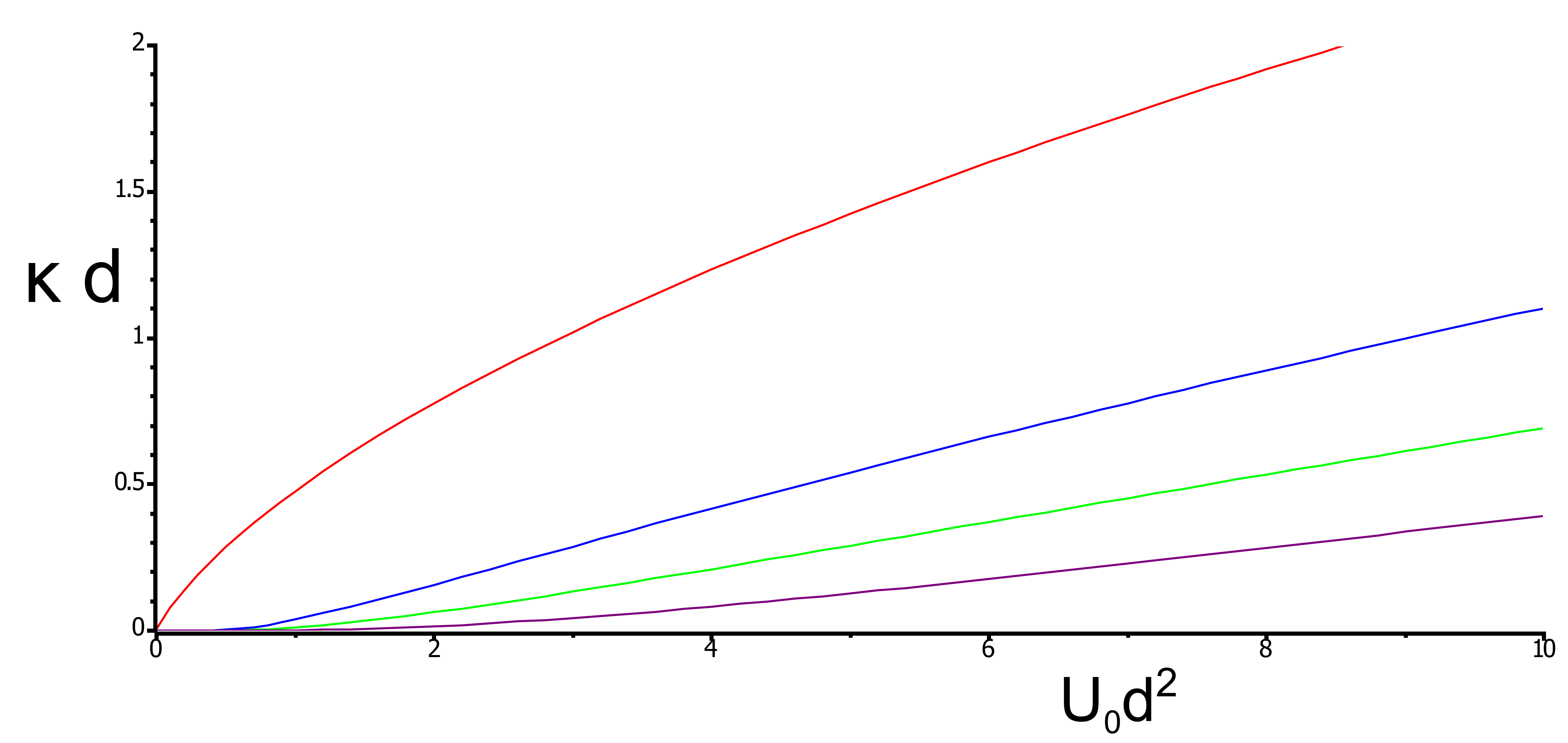}  
\caption{The progression of the four lowest eigenstates of the $p=0$ potential with system parameters $U_0 d^2$, found via solutions of Eq.~\eqref{p04} and Eq.~\eqref{p05}. }
\label{fig:p0energy}
\end{figure}

We plot in Fig.~\ref{fig:p0plots} the four lowest eigenstates respectively, detailing the usual alternating symmetric and antisymmetric solutions. The advance of these four states with the system parameters $U_0$ and $d$ is shown in Fig.~\ref{fig:p0energy}. 

\section{\label{p=1}Double-well: \lowercase{p}=1}

Considering a double hump potential profile, we set $p=1$ in Eq.~\eqref{SE2}. Eliminating the independent variable via the transformation $\xi = 2 \kappa d (1 \pm x/d)$ we intermediately obtain Whittaker's differential equation\cite{Gradshteyn}
\begin{equation}
\label{p10}
\psi''(\xi) + \left( -\frac{1}{4} + \frac{\mu}{\xi} + \frac{\frac{1}{4}-\nu^2}{\xi^2} \right) \psi(\xi) = 0.
\end{equation}
where
 \begin{equation}
\label{p11}
 \mu = \frac{U_0 d}{2 \kappa}, \qquad \nu  = \sqrt{\frac{1}{4} + U_0 d^2}.
\end{equation}
The square-integrable solution we desire is the Whittaker function of second kind, which can be constructed as follows\cite{Gradshteyn}
\begin{equation}
\label{p12}
 W_{\mu, \nu} (\xi) = \frac{\Gamma(-2\nu)}{\Gamma(\frac{1}{2}-\nu-\mu)} M_{\mu, \nu} (\xi) +
						\frac{\Gamma(2\nu)}{\Gamma(\frac{1}{2}+\nu-\mu)} M_{\mu, -\nu} (\xi),
\end{equation}
in terms of the Whittaker function of first kind, $M_{\mu, \nu} (\xi) = \xi^{\nu+\frac{1}{2}} e^{-\frac{\xi}{2}} _1F_1\left(\frac{1}{2}+\nu-\mu; 2\nu+1; \xi \right)$, where $_1F_1(\alpha; \beta; z)$ is a confluent hypergeometric function and $\Gamma(z)$ a gamma function. Eq.~\eqref{p12} decays for large $|\xi|$ as follows
\begin{equation}
\label{p12b}
 W_{\mu, \nu} (\xi) \sim e^{-\frac{\xi}{2}} \xi^{\mu},
\end{equation}
and is thus indeed a solution corresponding to a bound-state wavefunction. 

\begin{figure}[htbp] 
\includegraphics[width=0.45\textwidth]{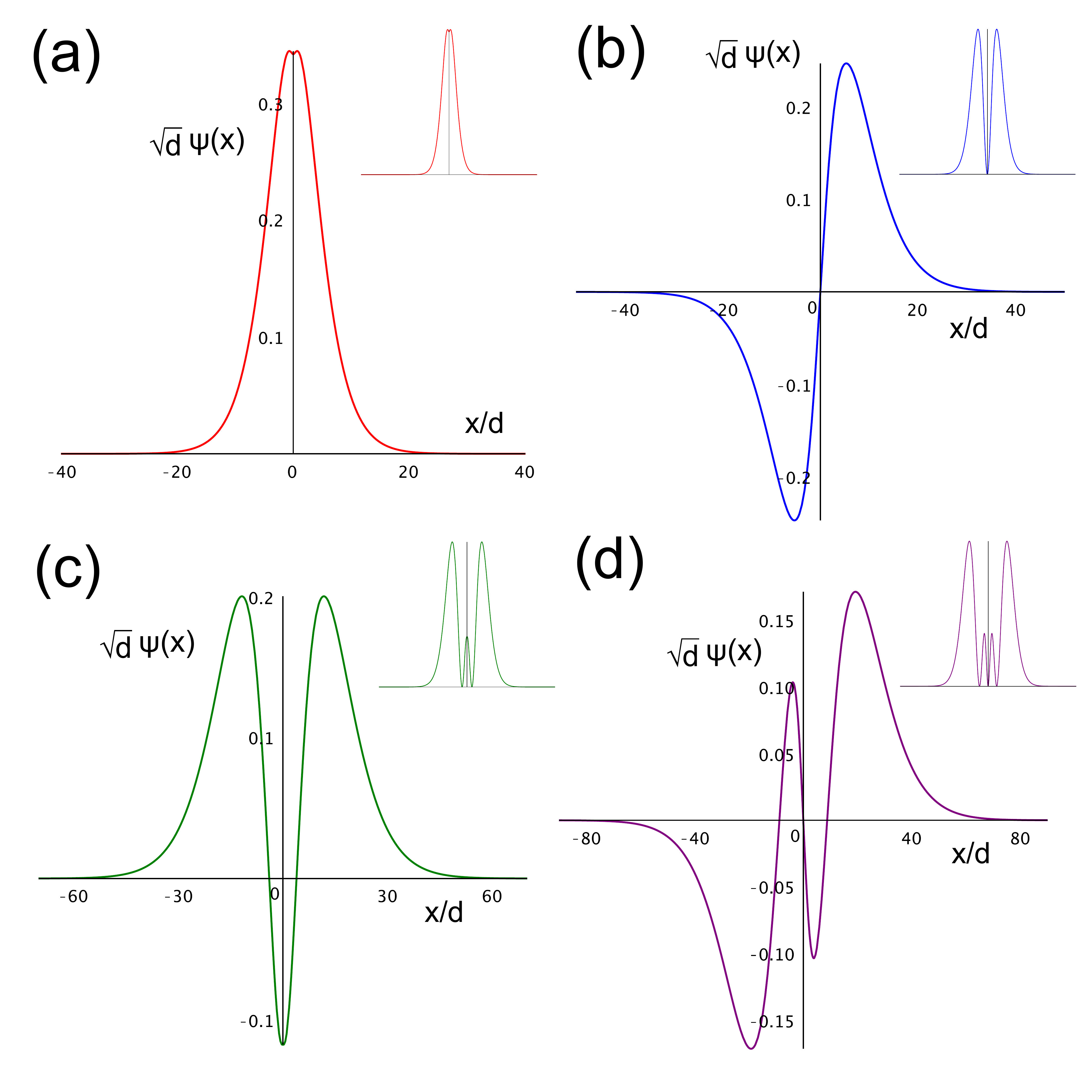}  
\caption{The four lowest eigenstates of the  $p=1$ potential, with $U_0 d^2 = 1$ and (a) $\kappa d = 0.408$, (b) $\kappa d = 0.290$, (c) $\kappa d = 0.222$, and (d) $\kappa d = 0.183$. Inset: probability density.}
\label{fig:p1plots}
\end{figure}

Upon ensuring both the full solution, 
\begin{equation}
\label{p12c}
 \psi_{\pm}(x) = \frac{c_{\pm}}{\sqrt{d}} W_{\mu, \nu} (2 \kappa d (1 \pm x/d)),
\end{equation}
and its first spatial derivative are continuous across the interface at the origin, we find the eigenvalues of antisymmetric solutions arise via 
\begin{equation}
\label{p13}
 W_{\mu, \nu} (2 \kappa d) = 0,
\end{equation}
which is coupled to the constraint $c_{+} = - c_{-}$. The symmetric solutions have eigenvalues governed by
\begin{equation}
\label{p14}
 W_{\mu, \nu}'(\xi) \bigg|_{x=0} = 0,
\end{equation}
along with the condition $c_{+} = c_{-}$. Eqs.~(\ref{p13},\,\ref{p14}) can be solved by the usual graphical or numerical methods and the remaining normalization constant  $c_{+}$ is found by square-integrating over the interval $(-\infty, \infty)$.

\begin{figure}[htbp] 
\includegraphics[width=0.45\textwidth]{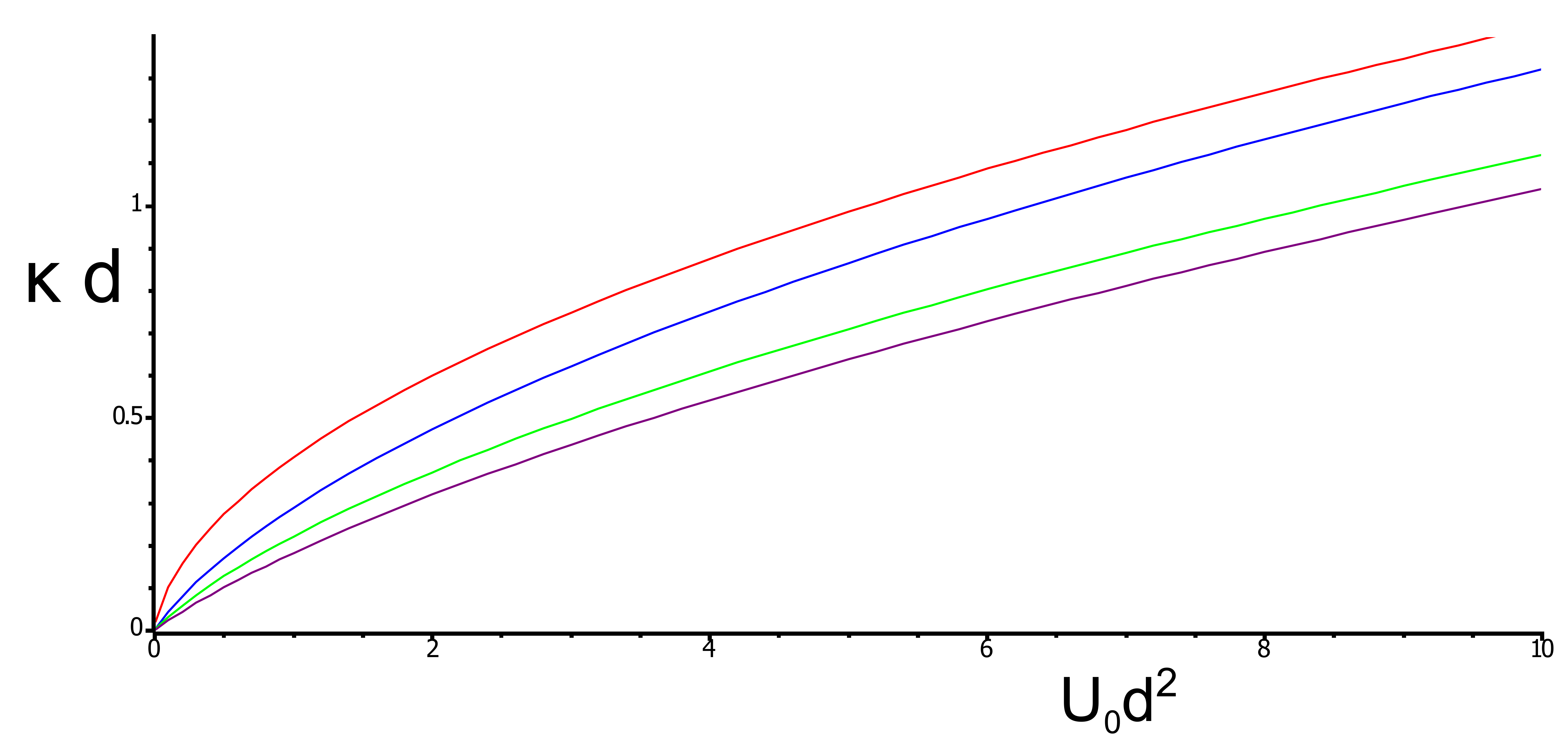}  
\caption{The progression of the four lowest eigenstates of the $p=1$ potential with system parameters $U_0 d^2$, found via solutions of Eq.~\eqref{p13} and Eq.~\eqref{p14}. }
\label{fig:p1energy}
\end{figure}

In Fig.~\ref{fig:p1plots} we display in our wavefunction plots the parity interchange of successive states, from the ground to the next three highest states, as expected for an even potential. The evolution of these states with modulation of the system parameters $U_0$ and $d$ is shown Fig.~\ref{fig:p1energy}.

\section{\label{p=2}Shallow single well: \lowercase{p}=2}

When taking the $p=2$ case, we shift where we measure the reference level and redefined constants, preferring to instead consider 

\begin{equation}
\label{p20}
 V(x) = -\frac{\hbar^2 }{2m} U_0 \left( 1 - \frac{ (|x|/d)^2}{(1+|x|/d)^2} \right).
\end{equation}

\begin{figure}[htbp] 
\includegraphics[width=0.45\textwidth]{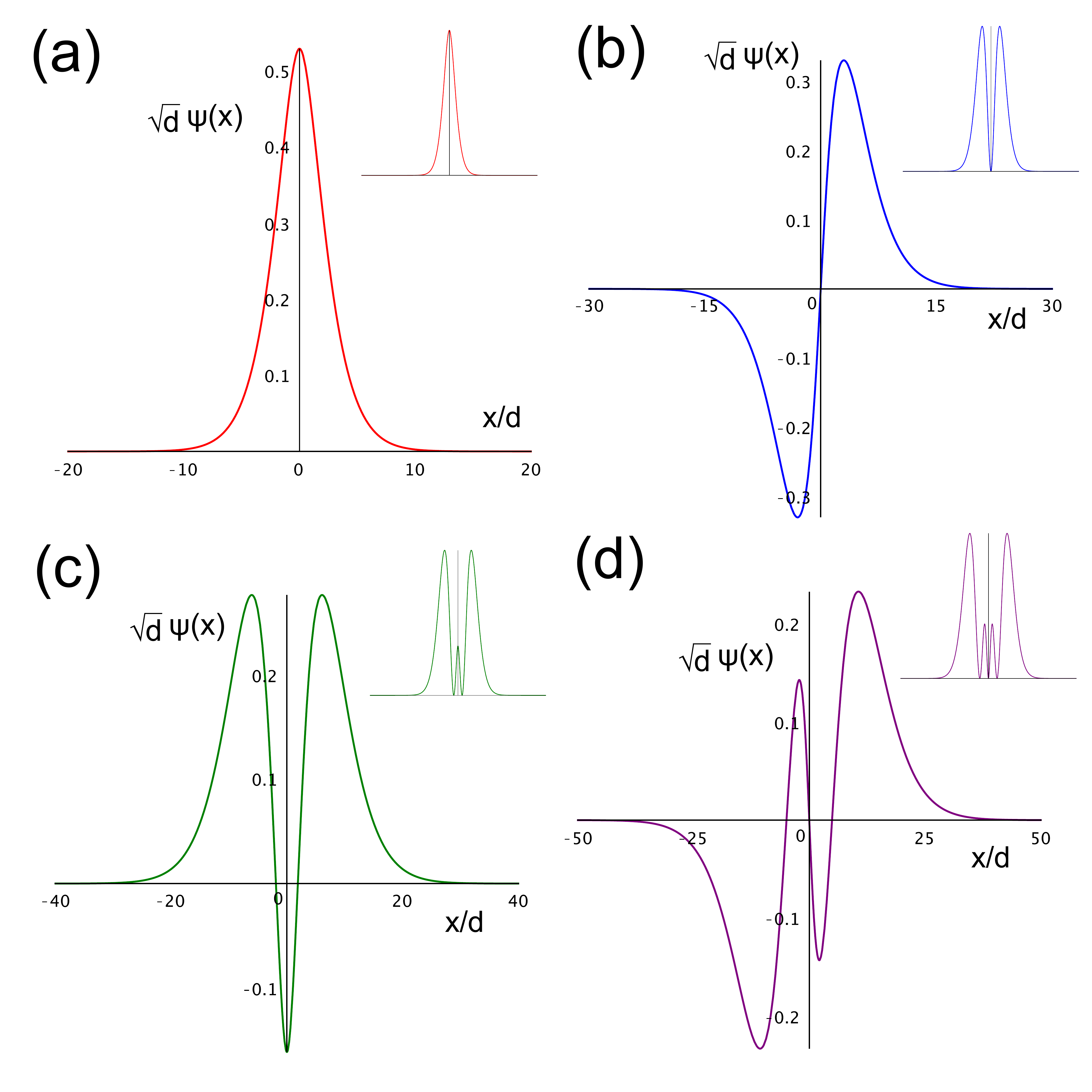}  
\caption{The four lowest eigenstates of the  $p=2$ potential, with $U_0 d^2 = 1$ and (a) $\kappa d = 0.796$, (b) $\kappa d = 0.532$, (c) $\kappa d = 0.425$, and (d) $\kappa d = 0.345$. Inset: probability density.}
\label{fig:p2plots}
\end{figure}

\begin{figure}[htbp] 
\includegraphics[width=0.45\textwidth]{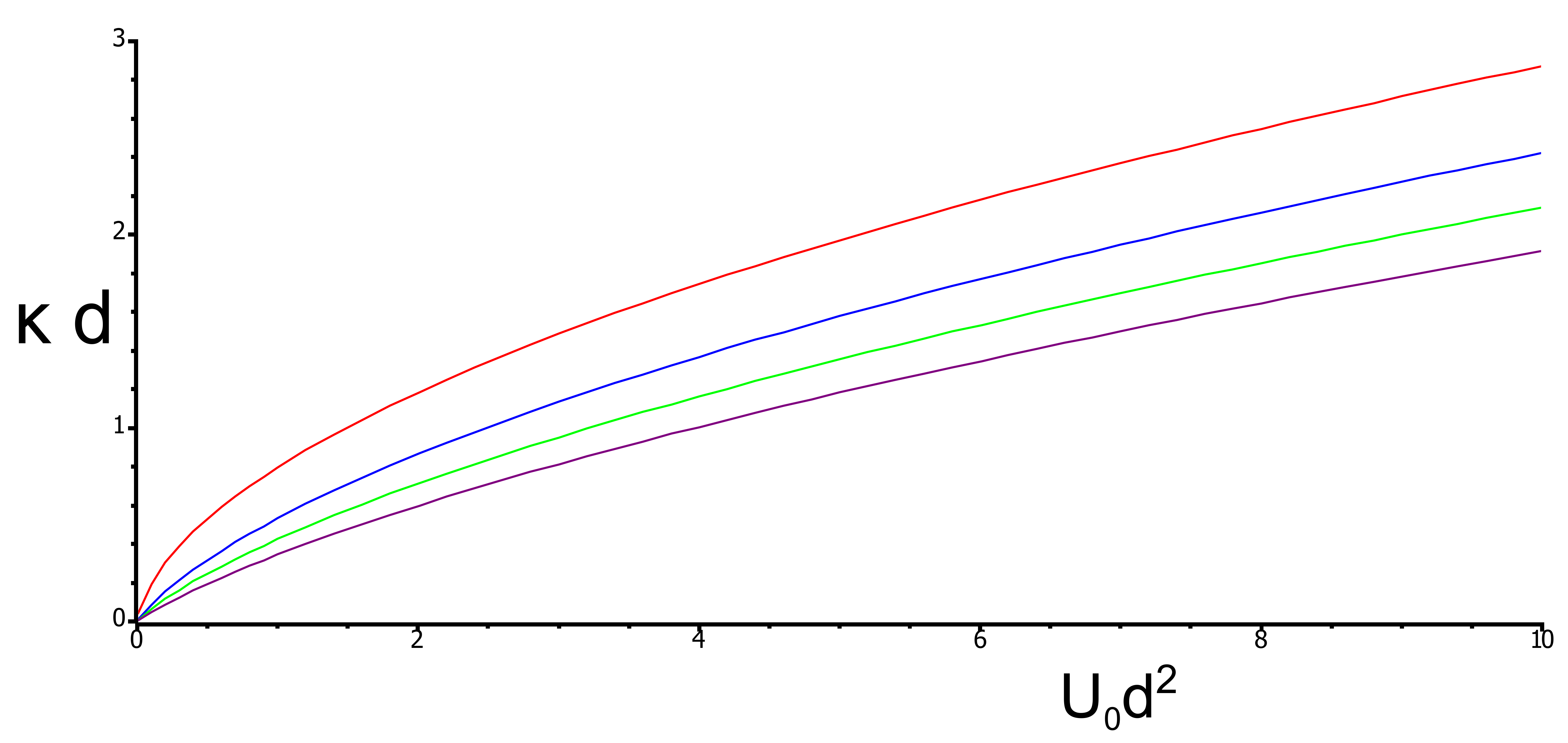}  
\caption{The progression of the four lowest eigenstates of the $p=2$ potential with system parameters $U_0 d^2$. }
\label{fig:p2energy}
\end{figure}

As in Sec.~\ref{p=1}, working in the variable $\xi = 2 \kappa d (1 \pm x/d)$ leads to a Whittaker differential equation, but now with a transformed $\mu$ parameter
 \begin{equation}
\label{p21}
 \mu \to \frac{U_0 d}{\kappa}.
\end{equation}
Remarkably, the $p=1$ solution  Eq.~\eqref{p12c}, along with both eigenvalue conditions Eqs.~(\ref{p13},\,\ref{p14}), solve this toy model problem upon making the above transformation, Eq.~\eqref{p21}. Fig.~\ref{fig:p2plots} and Fig.~\ref{fig:p2energy} show the behavior of example eigenstates and eigenvalues respectively. 

Differences between this shallower single well Eq.~\eqref{p20}, compared to the steeper well of Sec.~\ref{p=0}, is most noticeable in both the inter-energy-level separation, which is smaller, and the actual values of the energy levels, which are deeper (both features are measured as a function of the ground state energy). This is simply due to the wavefunctions, in going from a steeper to a shallower well, become (relatively) extended, squeezing and lowering the energy levels of the system.

\section{\label{conc}Discussion}

The structure of the Schr\"{o}dinger equation means the potential class investigated here with Eq.~\eqref{potential} is exactly-solvable in terms of Whittaker functions, and in fact remains so even with an additional Loudon-type\cite{Loudon} function also added 

\begin{equation}
\label{potential2}
V(x) = -\frac{\hbar^2}{2m} \left( U_0 \frac{ (|x|/d)^p}{(1+|x|/d)^2} + U_1 \frac{ (|x|/d)^q}{(1+|x|/d)} \right),
\end{equation}
\begin{equation*}
\label{potential3}
p = 0, 1, 2, \qquad q = 0, 1,
\end{equation*}

with the change of variable used in this work, and with only the parameters $\mu, \nu$ of the Whittaker functions being modified.

In conclusion, bound states solutions for a class of confining potential have been obtained with the one-dimensional Schr\"{o}dinger equation. In constructing our solutions, we made significant use of Whittaker functions\cite{note} to express the eigenstates, and found brief transcendental equations specify the allowed eigenvalues. Notably, the model systems include examples of both single wells and a double-well, which increases the variety of potential applications in physics and chemistry. We hope that these exact solutions will be of use in the construction of new physical models.

\section*{Acknowledgments}
We would like to thank M. Portnoi and N. Tufnel for useful discussions and D. St. Hubbins and D. Smalls for a critical reading of the manuscript. This work was supported by the EPSRC.



\begin{thebibliography}{100}
\bibitem{note}Markedly, the analysis in Sec.~\ref{p=0} could also be carried out with Whittaker functions, after using the identity $\xi^{1/2}  K_{\alpha} (\xi) = \left( \frac{\pi}{2}\right)^{1/2} W_{0, \alpha}(2 \xi)$.
\end{thebibliography}
\end{document}